\documentclass[twocolumn,showpacs,superscriptaddress,preprintnumbers,amsmath,amssymb,prl]{revtex4}
\usepackage[dvips]{graphicx,color}
\usepackage{graphicx}
\usepackage{dcolumn}
\usepackage{bm}
\begin{document}
\preprint{cond-mat}
\title{Compensation of Effective Field in the Field-Induced Superconductor 
$\kappa$-(BETS)$_2$FeBr$_4$ Observed by $^{77}$Se NMR}
\author{S.\ Fujiyama}
\email{fujiyama@ap.t.u-tokyo.ac.jp} 
\affiliation{Department of Applied Physics, University of Tokyo, Tokyo 113-8656, Japan}
\affiliation{CREST-JST, Kawaguchi, Saitama 332-0012, Japan}
\author{M.\ Takigawa}
\affiliation{Institute for Solid State Physics, University of Tokyo, Kashiwa, Chiba 277-8581, Japan}
\author{J.\ Kikuchi}
\thanks{Present address: Dept. of Physics, Meiji University, Kawasaki 214-8571, Japan}
\affiliation{Institute for Solid State Physics, University of Tokyo, Kashiwa, Chiba 277-8581, Japan}
\author{H-B.\ Cui}
\author{H.\ Fujiwara}
\thanks{Present address: Dept. of Chemistry, Osaka Prefecture University, Osaka 599-8570, Japan.}
\author{H.\ Kobayashi}
\affiliation{Institute for Molecular Science, Okazaki, Aichi 444-8585, Japan}
\affiliation{CREST-JST, Kawaguchi, Saitama 332-0012, Japan}
\date{\today}
\begin{abstract}
We report results of $^{77}$Se NMR frequency shift in the 
normal state of the organic charge-transfer-salt 
$\kappa$-(BETS)$_2$FeBr$_4$ which shows magnetic field-induced 
superconductivity (FISC). From a simple mean field
analysis, we determined the field and the temperature dependences 
of the magnetization $m_{\pi}$ of the $\pi$ conduction electrons on 
BETS molecules. We found that the Fe spins are antiferromagnetically 
coupled to the $\pi$ electrons and determined the exchange field to be 
$J$ = -2.3~T/$\mu_B$.  The exchange field from the fully saturated 
Fe moments (5~$\mu_B$) is compensated by an external field of 12~T.  This 
is close to the central field of the FISC phase, consistent with the 
Jaccarino-Peter local field-compensation mechanism for FISC (Phys. Rev. Lett. \textbf{9}, 290 (1962)).
\end{abstract}

\maketitle
There has been considerable interest in the correlation 
between conduction electrons and local spins in solid 
state materials. Competition between magnetic ordering 
due to the Ruderman-Kittel-Kasuya-Yosida (RKKY) interaction and the 
formation of Kondo singlet state accompanied by mass enhancement 
of conduction electrons is a canonical problem~\cite{Tsunetsugu1997}.
In the last decades, several organic charge-transfer-salts with 
anions containing magnetic ions have been synthesized~\cite{Day1992,Coronado2000}. 
It is widely recognized that physical properties of organic conductors 
are determined by a single band formed by the HOMO (highest occupied 
molecular orbitals) consisting of hybridized $\pi$-orbitals on donor 
molecules, and simple tight binding approximation works extremely 
well for the description of the HOMO band. Such simplicity is a 
significant advantage in the study of correlation between conduction 
$\pi$-electrons and local moments.

Recently, field-induced superconductivity (FISC) has been 
discovered in $\lambda$-(BETS)$_2$FeCl$_4$ under magnetic 
fields between 17~T and 40~T~\cite{Uji2001,Balicas2001}, 
where BETS stands for bis(ethylenedithio)tetraselenafulvalene. 
The FISC occurs only when the external field is applied parallel 
to the two dimensional conducting layers for the $\pi$-electrons.  
The FISC is rapidly suppressed by tilting the field, indicating 
onset of the orbital pair breaking. 

The FISC in $\lambda$-(BETS)$_2$FeCl$_4$ has been proposed to 
be due to the Jaccarino-Peter (JP) mechanism~\cite{Jaccarino1962,Fischer1972} 
originating from antiferromagnetic coupling between the Fe 
local moments and the $\pi$ electrons ($\pi$-$d$ interaction).  
Superconductivity is expected when the external field 
$H_\textrm{ext}$ nearly cancels 
the exchange field from Fe moments, minimizing the spin-Zeeman 
pair breaking effect. Indeed, the superconducting transition 
temperature $T_{c}$ in $\lambda$-(BETS)$_2$FeCl$_4$ shows a 
maximum at $H_\textrm{ext}$=32~T~\cite{Balicas2001}, which 
agrees well with the exchange field estimated from the Zeeman 
splitting of the Shubnikov-de Haas oscillation~\cite{Cepas2002,Uji2001B}. 
Also a non-magnetic analog $\lambda$-(BETS)$_2$GaCl$_4$ shows 
superconductivity at zero field.~\cite{HKobayashi1993}

\begin{figure}[b]
\centering
\includegraphics*[width=6cm]{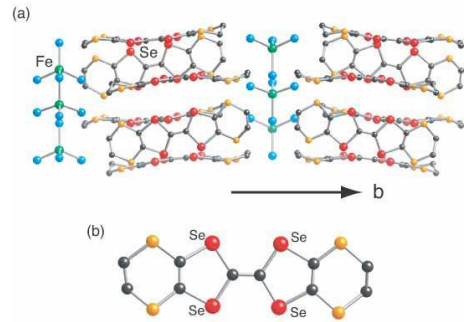}
\caption{(a) Crystal structure of $\kappa$-(BETS)$_2$FeBr$_4$. 
(b) A BETS donor molecule with Se atoms.}
\label{fig:Xtal}
\end{figure}

A $\kappa$-type polymorph of BETS based material, 
$\kappa$-(BETS)$_2$FeBr$_4$ has been also reported to 
show FISC between 10 T and 15 T with the maximum of 
$T_{c}$=0.3~K at 12.5~T~\cite{Konoike2004,Fujiwara2002}. Similar to the 
$\lambda$-type polymorph, two dimensional conducting BETS 
layers parallel to the $ac$-plane alternate with the 
anion layers as shown in Fig.~\ref{fig:Xtal}. Under zero magnetic 
field, antiferromagnetic order of Fe moments occurs below 
$T_N$= 2.5 K where the system remains metallic and superconductivity 
appears below $T_\textrm{c}$ = 1.1 K. This is in contrast to 
the case of $\lambda$-(BETS)$_2$FeCl$_4$, where antiferromagnetic order
of Fe spins at zero field at $T_{N}$=8~K causes a metal-insulator 
transition.~\cite{Kobayashi1996} 

Nuclear magnetic resonance (NMR) is an excellent probe to 
measure local spin density, therefore, can be used to test 
the validity of the JP mechanism for FISC. Measurements of 
Mo NMR frequency shift was reported in a Shevrel phase 
intermetallic FISC compound (Eu,Sn)Mo$_6$(S,Se)$_8$~\cite{Fradin1977}, 
for which the JP mechanism was proposed between conduction electrons 
and doped Eu spins~\cite{Fischer1975,Meul1984}. Although the results  
support partial cancellation of external field by the exchange 
field from the Eu spins, quantitative analysis was 
hindered by complication of the states near the Fermi level 
with both the $s$ and the $d$ bands.  

In this Letter, we report the frequency shift of $^{77}$Se NMR in 
$\kappa$-(BETS)$_2$FeBr$_4$, both as a function of $H_\textrm{ext}$
at a constant temperature (1.5~K) and as a function of temperature 
at a constant field (15.5~T). The results shows that the spin polarization of 
$\pi$ electrons is determined as the product of effective field, which is the 
sum of $H_\textrm{ext}$ and the antiferromagnetic exchange field from Fe spins,
and the local susceptibility of $\pi$ electrons, which is independent 
of magnetization of Fe spins, $m_{d}$.  The exchange field is estimated to be 
in the range 10 - 12~T for the saturated Fe magnetization of 5$\mu_{B}$/Fe.  This agrees well
with the field for the highest $T_{c}$ in the FISC phase, thus supporting the 
JP mechanism. 

The $^{77}$Se NMR experiments were performed using a rectangular 
plate-like single crystal of $\kappa$-(BETS)$_2$FeBr$_4$ 
grown by the standard electrochemical oxidation method. 
The magnetic susceptibility follows a Curie-Weiss law 
$\chi(T)=C/(T-\theta)$ with $\theta$=-5.5~K and 
$C$=4.70~K$\cdot$emu/mol.  The Curie constant 
is close to the value expected for the localized high spin 
state of Fe$^{3+}$ (4.4~K$\cdot$emu/mol for $S$=5/2 and 
$g$=2) and the negative value of $\theta$ indicates 
antiferromagnetic interactions~\cite{Fujiwara2001}.
A unit cell contains eight BETS molecules, which are 
related by symmetry operations of the crystal structure 
(space group $P_{nma}$). We consider the cases in which 
the field is directed along the $a$, $b$, or $c$-axis of the 
crystal, which are invariant under all symmetry operations. 
Then all BETS molecules should yield identical NMR spectrum.
We expect $^{77}$Se NMR spectrum composed of four distinct 
lines corresponding to the four inequivalent Se sites in a BETS 
molecule (Fig.~\ref{fig:Xtal}b). 

\begin{figure}[t]
\centering
\includegraphics*[width=8cm]{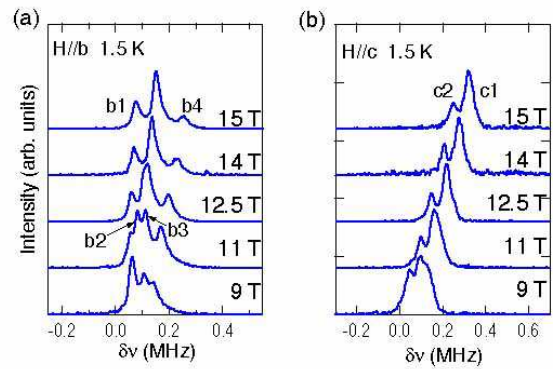}
\caption{$^{77}$Se NMR spectra at $T$=1.5~K for the external field  
along the $b$-axis (a) and along the $c$-axis (b). The resolved four (two) 
peaks are denoted as b1, b2, b3, and b4 (c1 and c2).}
\label{fig:Fdep}
\end{figure}

The $^{77}$Se NMR spectra at $T$=1.5~K obtained for several different 
values of the external field along the $b$- and $c$-axes are shown in 
Fig.~\ref{fig:Fdep}. The horizontal axis represents the frequency 
shift $\delta\nu = \nu - \gamma_N H_\textrm{ext}$, where $\nu$ 
is the signal frequency and $\gamma_N$=8.118~MHz/T 
is the nuclear gyromagnetic ratio. While four resonance lines are 
indeed observed for $H \parallel b$ (Fig.~\ref{fig:Fdep}a),  
the spectrum is broader and only two peaks are clearly 
resolved for $H \parallel c$ (Fig.~\ref{fig:Fdep}b). 

The hyperfine field acting on a Se nucleus comes from both the 
$\pi$ electrons and the Fe spins. Thus the frequency shift at 
the site $i$ ($i$=1 - 4) for the field along the $\alpha$-direction 
has two terms,
\begin{equation}
\delta\nu_{i, \alpha} = \gamma_N A^{\pi}_{i, \alpha} m_{\pi} + 
\gamma_N (A^\textrm{dip}_{i, \alpha} + B_{i} ) m_{d}.
\label{shift}
\end{equation}
The first term is the contribution from the magnetization 
of $\pi$ electrons ($m_{\pi}$) through the hyperfine coupling 
$A^{\pi}_{i, \alpha}$.  The second term is the contribution from 
the Fe magnetization $m_{d}$ through the direct dipolar coupling 
$A^\textrm{dip}_{i, \alpha}$ and the transferred hyperfine 
coupling $B_{i}$. While the coupling constant $A^{\pi}_{i, \alpha}$ 
is determined by the fractional weights of Se atomic orbitals 
participating in the HOMO of BETS molecules, $B_{i}$ is 
due to hybridization between the Fe-$d$ states and the Se atomic 
orbitals which do not participate in the HOMO. Since the relevant
states for $B_{i}$ are the inner-core states not the 
outer-most 4$p$ states, we expect $B_{i}$ to be nearly isotropic.   

\begin{figure}[t]
\centering
\includegraphics*[width=7cm]{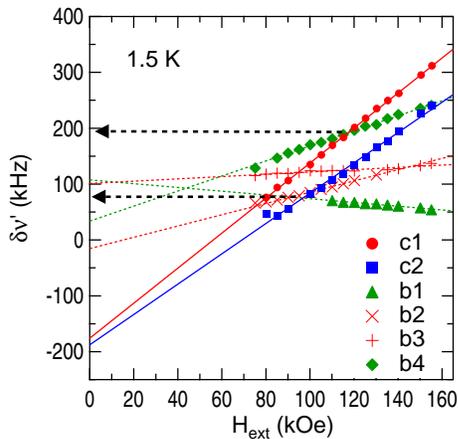}
\caption{$H_\textrm{ext}$ dependence of $\delta\nu^{\prime}$ at $T$=1.5~K. 
The solid (dotted) lines are the fits to straight lines for 
$H \parallel c$ ($H \parallel b$). The arrows show the values of  
$B_{i}m_{d}$.}
\label{fig:Fdep_Layout}
\end{figure}

The dipolar coupling tensors are nearly diagonal and calculated as ($A^\textrm{dip}_{i,a}, 
A^\textrm{dip}_{i,b}, A^\textrm{dip}_{i,c}$)=(-4.1, 7.3, -3.2), 
(-4.4, 7.8, -3.4), (-4.2, 8.7, -4.5), (-4.2, 8.5, -4.3)~10$^{-3}$T/$\mu_B$ 
for the four Se sites.  Although we have not succeeded in assigning resonance 
lines to specific sites, the dipolar field is more or less the 
same for all sites.  Taking the values 
$A^\textrm{dip}_{i,c}$=-0.004 and $A^\textrm{dip}_{i,b}$=0.0085~T/$\mu_B$ 
for all sites and subtracting them from the observed shift, 
we obtain $\delta\nu^{\prime}_{i, \alpha}=\gamma_N A^{\pi}_{i, \alpha} m_{\pi} + \gamma_N
B_{i}m_{d}$.  The $H_\textrm{ext}$ dependence of 
$\delta\nu^{\prime}$ at $T$=1.5~K is shown in Fig.~\ref{fig:Fdep_Layout}
for various sites and different field directions. Note that the Fe moments
are completely saturated ($m_{d}$=5~$\mu_{B}$) in this field range at 1.5~K.

In all cases, $\delta\nu^{\prime}$ show linear dependences on 
$H_\textrm{ext}$. Since the Fe moments are saturated and fluctuations
are negligible, we may consider the effects of $\pi$-$d$ exchange 
interaction in a mean-field approximation. The polarization of $\pi$ electrons 
is expresses as
\begin{equation}
m_{\pi}=\chi_{\pi} (H_\textrm{ext}+Jm_{d}),
\label{MeanField}
\end{equation}
where $Jm_{d}$ is the exchange field acting on the $\pi$-electrons from 
the Fe moments and $\chi_{\pi}$ is the local susceptibility of the $\pi$
electrons in the absence of $\pi$-$d$ exchange interaction.
Combining Eqs.~(\ref{shift}) and (\ref{MeanField}), the slope 
of the lines in Fig.~\ref{MeanField} gives the value of 
$A^{\pi}_{i, \alpha} \chi_{\pi}$, which is listed in the upper 
panel of Table~\ref{tab:J}. These values are within the range of
the anisotropic $^{77}$Se NMR shift reported in a non-magnetic 
analog $\kappa$-(BETS)$_2$GaCl$_4$ at low temperatures, which is 
between -0.1~\% and 0.6~\%~\cite{Takagi2003}.  This supports 
the validity of our mean field definition of $\chi_{\pi}$. 
\begin{table}[b]
\caption{\label{tab:J}Estimated values of 
$A_{i, \alpha}^{\pi}\chi_{\pi}$ and 
$J$ by the $H_\textrm{ext}$ (top) and $T$ (bottom) 
dependences of the frequency shifts.}
\begin{tabular}{crrrrrrr}
lines &c1& c2 &c2'&b1&b2&b3&b4\\
\hline
$A_{i, \alpha}^{\pi}\chi_{\pi}$[\%]& 0.39 & 0.35& &-0.04&0.12&0.02&0.17 \\
$J$[T/$\mu_B$]& -2.33&-2.03&&-1.93&-2.04&-2.35&-2.32\\
\hline
$A_{i, \alpha}^{\pi}\chi_{\pi}$[\%]& 0.53 & 0.43& 0.44&-0.05&0.14&0.004&0.17 \\
$J$[T/$\mu_B$]& -2.48&-2.19&-1.83&-2.39&-2.25&-1.88&-2.16\\
\end{tabular}
\end{table}

If the JP field-compensation mechanism is valid, we expect 
$m_{\pi}$ to become zero at the central field of the FISC
phase. Since $B_{i}$ is assumed to be isotropic, 
$\delta\nu^{\prime}_{i, \alpha}$ is expected to be 
independent of the field direction when $m_{\pi}$=0. Thus 
the two lines for $H \parallel c$ and $H \parallel b$ 
in Fig.~\ref{fig:Fdep_Layout} corresponding to the 
same sites should cross at a field where $m_{\pi}$=0. 
By inspecting Fig.~\ref{fig:Fdep_Layout}, we indeed recognize 
that each line for $H \parallel b$ crosses one line 
for $H \parallel c$ in the range 10 - 12~T. 
This field range agrees approximately with the 
field for the highest $T_{c}$ in the FISC phase (12.5~T). 

Unfortunately, we could not follow evolution of the 
NMR peak frequencies as the field is rotated in the 
$bc$-plane due to broad line-width and additional line 
splitting for the field away from the $b$ or $c$ direction.  
Therefore, we have to make a best guess to assign each line for 
$H \parallel b$ to a specific line for $H \parallel b$.
Since there are only two peaks for $H \parallel c$, at least one
of these corresponds to more than one site. Crossing of three 
lines (c2, b1, b2) at nearly the same field (9.7~T and 10.0~T) 
strongly suggests that the b1 and the b2 peaks evolve into the c2 peak
upon field rotation. We then obtain $\gamma_N B m_{d}$
=85~kHz for these sites. Similarly, the b4 and the c1 lines crossing 
at 11.8~T are assigned to a common 
site with $\gamma_N B m_{d}$=190~kHz. There is small ambiguity 
about the b3 line, which crosses the c1 line at 
9.5~T and the c2 line at 11.5~T. However, both cases lead to a similar 
value of $\gamma_N B m_{d}$=120~kHz because of very small 
slope of the b3 line.   

\begin{figure}[t]
\centering
\includegraphics*[width=7cm]{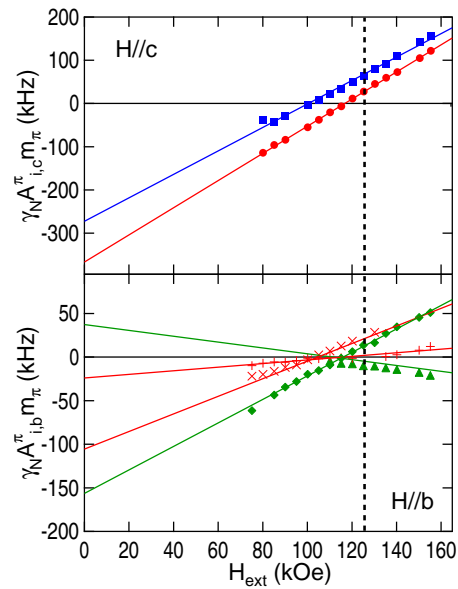}
\caption{Frequency shifts due to magnetization 
of $\pi$ electrons at 1.5~K.}
\label{fig:HFDDD}
\end{figure}

By subtracting $\gamma_N B_{i} m_{d}$ from 
$\delta\nu^{\prime}_{i, \alpha}$, we obtain
$\gamma_N A^{\pi}_{i, \alpha} m_{\pi}$ and plotted in 
Fig.~\ref{fig:HFDDD}.  The compensation fields 
$H_\textrm{ext}= - Jm_{d}$ obtained from the condition
$m_{\pi}$=0 are in the range 10 - 12~T, in agreement with 
the center of the FISC phase. Thus our results support the JP mechanism. 
The values of $J$ are listed in the upper panel of Table~\ref{tab:J}. 
The estimated values of $J$ are \textit{(1) nearly the same for all 
four Se sites} and \textit{(2) independent of the 
directions of $H_{\rm ext}$.} A small distribution 
in the estimated value of $J$ could be due to 
insufficient spectral resolution for $H \parallel c$ 
and/or possible small anisotropy of $B_{i}$.  Our microscopic estimate
of the compensation field agrees well with the earlier estimate
of the exchange field (12~T) based on the analysis of the Schubnikov-de Haas 
oscillations~\cite{Cepas2002,Uji2001C}. 

\begin{figure}[t]
\centering
\includegraphics*[width=7cm]{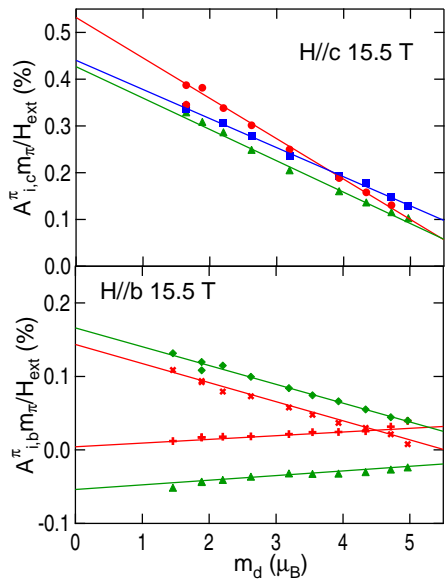}
\caption{Frequency shifts due to the magnetization of $\pi$ electrons 
are plotted against the magnetization of Fe moments $m_{d} \equiv
B_{5/2}(5\mu_{B}H_\textrm{ext}/k_B T)$ at $H_\textrm{ext}$=15.5~T.}
\label{fig:HTdepKM}
\end{figure}

An independent support for the validity of our mean-field
analysis was obtained from the temperature dependence of 
the shift measured at a constant field of 15.5~T.  At this field,  
$m_{d}$ is well represented by the Brillouin function 
$B_{5/2}(5\mu_{B}H_\textrm{ext}/k_B T)$.
By subtracting $\gamma_{N} (A^\textrm{dip}_{i, \alpha} + B_{i}) m_{d}$ 
from the measured shift, we obtained the shifts due to $m_{\pi}$ 
($A^{\pi}_{i, \alpha} m_{\pi}/H_\textrm{ext}$), which is plotted against 
$m_{d}$ in Fig.~\ref{fig:HTdepKM}. The plot yields a straight
line again for all sites and field directions. If we assume no 
$T$-dependence for $\chi_{\pi}$ as supported by nearly 
$T$-independent NMR shift in $\kappa$-(BETS)$_2$GaCl$_4$~\cite{Takagi2003}, 
the slope and the intercept at $m_{d}$=0 give the values of $
A^{\pi}_{i, \alpha} \chi_{\pi} J$ and $A^{\pi}_{i, \alpha} \chi_{\pi}$, 
respectively.  The values of $A^{\pi}_{i, \alpha} \chi_{\pi}$ 
and $J$ thus obtained are listed in the lower panel of Table~\ref{tab:J}.  
The independent sets of data for the $H_\textrm{ext}$-dependence and 
the $T$-dependence of the shift lead to similar values of the 
parameters. This provides a strong support for the validity of the 
mean-field analysis of Eqs.~\ref{shift} and \ref{MeanField}. 

The exchange $\pi$-$d$ interaction of this material is estimated theoretically within the second order purturbation by use of the extended H\"{u}ckel calculation. 
When $d$ spins are polarized along the field direction, the $\pi$-$d$ interactions works simply on each molecule as an exchange field as $H=\Sigma_j {\cal J}_j \langle S \rangle$ by neglecting the spin fluctuation. Here ${\cal J}_j$ is $\pi$-$d$ interaction between molecules and surrounding $d$ spins with expectation value $\langle S \rangle$. The theoretical estimation amounts to ${\cal J}\equiv \Sigma_j {\cal J}_j/g\mu_B$=-2.25 (Hotta~\cite{Hotta2000}) or -3.0~T/$\mu_B$ (Mori~\cite{Mori2002}), which agrees very well with our estimate $J$=-2.3~T/$\mu_B$.

For a system with strong $\pi$-$d$ hybridization leading
to local screening of $d$ spins by conduction electrons, a gap 
may be developed in the spin excitation spectrum of the 
$\pi$ electrons and strong $T$ dependence of 
$\chi_{\pi}$ would be expected.~\cite{Tsunetsugu1997} In such a case, 
we would expect a deviation from the linear relation between $m_{\pi}$ and
$m_{d}$, which is not observed (Fig.~\ref{fig:HTdepKM}). 
We also expect significant discrepancy in the estimated
values of $J$ determined from the $H_\textrm{ext}$ dependence and 
the $T$ dependence of the shifts, again in contradiction 
to the observation. We therefore conclude that $\chi_{\pi}$ is nearly 
independent of temperature showing Pauli paramagnetic 
character above 1.5 K.

It should be noted that we are interested mainly in the behavior at 
high magnetic field, where the Zeeman energy for the large 
$S$=5/2 Fe spins is much greater than the $\pi$-$d$ exchange 
interaction, $-(g^2 \mu_B/k_B)J$=6~K.  Hence we do not expect quantum fluctuations 
or the Kondo screening to be important.  This is consistent with no 
indication of spin-gap behavior 
in $\chi_{\pi}$ above 1.5 K.

In summary, the frequency shift of $^{77}$Se NMR in the 
field-induced superconductor $\kappa$-(BETS)$_2$FeBr$_4$ was found 
to depend linearly on $H_\textrm{ext}$ at $T$=1.5~K and the magnetization of 
Fe spins ($m_{d}$) at $H_\textrm{ext}$=15.5~T. The results are 
consistent with the molecular field approximation for the $\pi$-$d$ exchange interaction. 
The local susceptibility of the $\pi$ electrons $\chi_{\pi}$ is 
independent of $m_{d}$ and shows Pauli-paramagnetic behavior above 1.5 K.
The exchange field on the $\pi$ electrons from Fe moments is canceled 
near 12~T where the FISC occurs. We have determined microscopically the antiferromagnetic 
$\pi$-$d$ interaction $J$=-2.3~T/$\mu_B$, that is consistent with the 
mechanism of local field compensation for FISC first proposed by Jaccarino and Peter. 

We acknowledge K.\ Kanoda, C.\ Hotta, T.\ Mori and E.\ Fujiwara for fruitful discussions. 
This work was supported by the JSPS (No.\ 17740213) and the MEXT 
(No.\ 13NP0201, 13640375, and 16076204). 

\begin{acknowledgments}
\end{acknowledgments}


\begin{thebibliography}{}
\bibitem{Tsunetsugu1997}H.\ Tsunetsugu, M.\ Sigrist, and K.\ Ueda, Rev.\ Mod.\ Phys.\ \textbf{69}, 809 (1997).
\bibitem{Day1992}P.\ Day, \textit{et al.}, J.\ Am.\ Chem.\ Soc.\ \textbf{114}, 10772 (1992), 
\bibitem{Coronado2000}E.\ Coronado \textit{et al.}, Nature \textbf{408}, 447 (2000).
\bibitem{Uji2001}S.\ Uji \textit{et al.}, Nature \textbf{410}, 908 (2001).
\bibitem{Balicas2001}L.\ Balicas \textit{et al.}, Phys.\ Rev.\ Lett.\ \textbf{87}, 067002 (2001).
\bibitem{Jaccarino1962}V.\ Jaccarino and M.\ Peter, Phys.\ Rev.\ Lett.\ \textbf{9}, 290 (1962).
\bibitem{Fischer1972}\O.\ Fischer, Helv.\ Phys.\ Acta, \textbf{45}, 331 (1972).
\bibitem{Cepas2002}O.\ C\'{e}pas, R.H.\ McKenzie, and J.\ Merino, Phys.\ Rev.\ B \textbf{65}, 100502 (2002).
\bibitem{Uji2001B}S.\ Uji \textit{et al.}, Phys.\ Rev.\ B\textbf{64}, 024531 (2001).
\bibitem{HKobayashi1993}H.\ Kobayashi \textit{et al.}, Chem.\ Lett.\ 1559 (1993).
\bibitem{Konoike2004}T.\ Konoike \textit{et al.}, Phys.\ Rev.\ B \textbf{70}, 094514 (2004).
\bibitem{Fujiwara2002}H.\ Fujiwara \textit{et al.}, J.\ Am.\ Chem.\ Soc.\ \textbf{124}, 6816 (2002).
\bibitem{Kobayashi1996}H.\ Kobayashi \textit{et al.}, J.\ Am.\ Chem.\ Soc.\ \textbf{118}, 368 (1996).
\bibitem{Fradin1977}F.Y.\ Fradin \textit{et al.}, Phys.\ Rev.\ Lett.\ \textbf{38}, 719 (1977).
\bibitem{Meul1984}H.W.\ Meul, C.\ Rossel, M.\ Decroux, and O.\ Fischer, Phys.\ Rev.\ Lett.\ \textbf{53}, 497 (1984).
\bibitem{Fischer1975}\O.\ Fischer \textit{et al.}, J.\ Phys.\ C \textbf{8}, L474 (1975).
\bibitem{Fujiwara2001}H.\ Fujiwara \textit{et al.}, J.\ Am.\ Chem.\ Soc.\ \textbf{123}, 306 (2001).
\bibitem{Takagi2003}S.\ Takagi \textit{et al.}, J.\ Phys.\ Soc.\ Jpn.\ \textbf{72}, 483 (2003).
\bibitem{Uji2001C}S.\ Uji \textit{et al.}, Physica B\textbf{298}, 557 (2001). 
\bibitem{Hotta2000}C.\ Hotta and H.\ Fukuyama, J.\ Phys.\ Soc.\ Jpn.\ \textbf{69}, 2577 (2000).
\bibitem{Mori2002}T.\ Mori and M.\ Katsuhara, J.\ Phys.\ Soc.\ Jpn.\ \textbf{71}, 826 (2002).
\end{thebibliography}
\end{document}